\documentclass[preprint,showpacs,amsmath,amstex,amssymb,mathfonts,prl]{revtex4-1}
\usepackage{graphicx}

\usepackage{tipa}
\usepackage{bm}

\usepackage{amsmath}
\usepackage{amssymb}
\usepackage{amsthm}
\usepackage{amsfonts}
\usepackage{float}

\begin{document}

\title{Microscopic Statistical Characterisation of the Congested Traffic Flow and Some Salient Empirical Features}

\author{Bo Yang, Ji Wei Yoon and Christopher Monterola}
\affiliation{Complex Systems Group, Institute of High Performance Computing, A*STAR, Singapore, 138632.}
\date{\today}

\date{\today}
\begin{abstract}
We present large scale and detailed analysis of the microscopic empirical data of the traffic flow, focusing on the non-linear interactions between the vehicles when the traffic is congested. By implementing a ``renormalisation" procedure that averages over relatively unimportant factors, we extract the effective dependence of the acceleration on the vehicle headway, velocity and relative velocity, that characterises not just a few vehicles but the traffic system as a whole. Several interesting features of the detailed vehicle-to-vehicle interactions are revealed, including the stochastic distribution of human responses, relative importance of non-linear terms in different density regimes, symmetric response to the relative velocity, and the insensitivity of the acceleration to the velocity within a certain headway and velocity range. The latter leads to a multitude of steady-states without a fundamental diagram, showing strong evidence of the ``synchronised phase", substantiated with additional microscopic details. We discuss the richness of the available data for the understanding of the human driving behaviours, and its usefulness in construction and calibration of both deterministic and stochastic microscopic traffic models.
\end{abstract}

\maketitle 
Modelling the traffic flow as an interacting physical system has been an active field of research for more than six decades\cite{kernerbook,treiberbook,reviewpaper,reviewpaper1}. The traffic system in general can be treated as a one-dimensional many-body systems with non-linear anisotropic interactions. The components in the system are \emph{non-identical}. This crucial difference from the more conventional many-body physical systems is usually circumvented by assuming the existence of a representative agent: the traffic system can be modelled by identical components governed by the same behaviour. Such ``model vehicle" and its equations of motion, if they exist, should also be obtained statistically in a rather non-trivial way. It is of fundamental interest both theoretically and empirically to understand what universal characteristics of the traffic system can be captured by such simplifications, and how the equations of motion can be generated from the empirical data.

Many phenomenological models with identical drivers have been proposed with simple and effective equations of motions that can successfully capture some of the most important features of the traffic flow. On the other hand, controversies still remain\cite{kernercrit,helbingcrit} on the nature of the dynamics of the traffic flow, especially when the flow is congested with emerging complex spatiotemporal patterns. The non-linear interactions, compounded with the near impossibility of controlling the actual traffic flow as an experimental system, make it highly nontrivial to compare the numerical simulations of the models with the empirical observations. 

Most popular traffic models are tuned with empirical driving behaviours by observing the dynamics of the vehicles either in an experimental setting or in the actual traffic flow\cite{shamoto,helbing,neubert,hamdar,tian,knospe}. The parameters in the models are usually tuned from a relatively small number of vehicles, and in general that is sufficient at least to estimate the right order of magnitude for the parameters. One should note, however, since the tuning is done for specific models, the fidelity of the resulting models depends on the assumptions built into the model constructions; certain detailed driving behaviours are already filtered by the models themselves. It is also important to realise that the dynamics of the a small number of individual vehicles even in the actual traffic flow may not characterise the interacting many-body traffic system well. The traffic system is completely defined by the distribution of the vehicle types, driver demographics and the road configuration, and a more thorough sampling may be needed especially for the construction of a representative ``model vehicle".

In this Letter, we present and analyse the data from the empirical traffic flows from a stretch of expressway in Singapore. Microscopic empirical data including the individual vehicle accelerations, headways, velocities and relative velocities are extracted from the videos of such traffic flow via image processing and machine learning, over a period of more than six months. In reality, the vehicle acceleration depends on the vehicle and driver types and many other (possibly stochastic) factors. While it is not possible to measure all the factors influencing the vehicles dynamics, we employ a ``renormalisation" approach to average over unimportant and rapidly fluctuating factors, to obtain an effective dependence of the acceleration on the important dynamical variables. This leads to a characteristic map of the acceleration $a$ as a function of the headway $h$, velocity $v$ and relative velocity $\Delta v$. The map reveals some interesting features of the detailed driving behaviours, including the speed synchronisation with certain range of the headway and velocity in the congested traffic. It can also be used for the construction of a ``master traffic model" and the validation of different types of the models being proposed in the literature.

\begin{figure}
\begin{center}
\includegraphics[height=9cm]{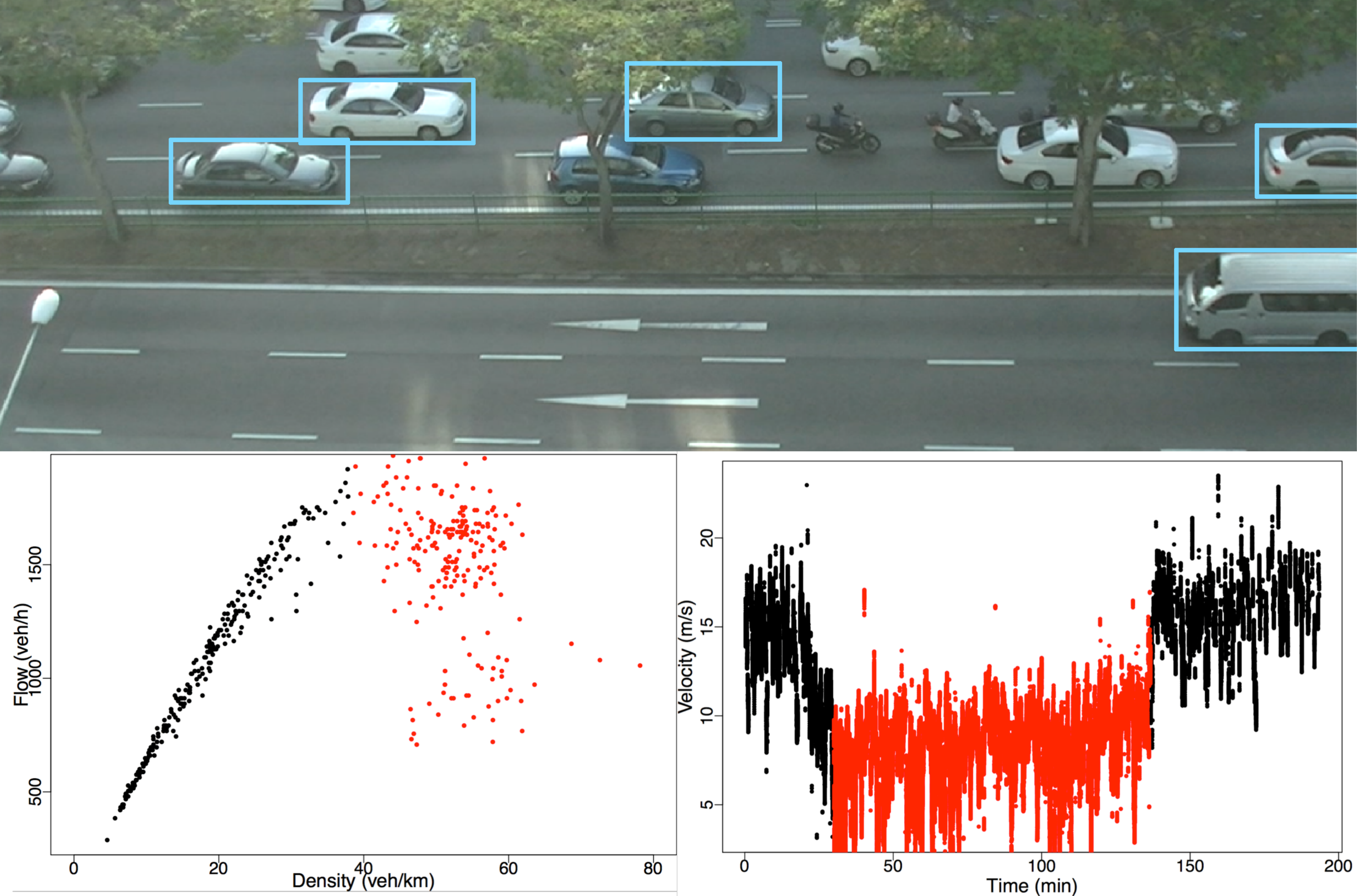}
\caption{The top figure is a snapshot of part of the traffic lanes included in the video analysis. The video is taken at 250 frames per second. Algorithms are implemented for accurate extraction of the positions, velocities and accelerations of the vehicles, excluding the effects of the shadows and trees blocking the view. The boxed vehicles are identified by our machine learning algorithm. The bottom left figure is the flow-density plot of the traffic flow of the fast lane traffic, measured over an entire week; the bottom right figure is the measured velocities of each vehicles as a function of the time. The red part of the plots come from the morning peak hours, during which the traffic is congested.}
\label{video}
\end{center}
\end{figure}

We first briefly introduce the methodologies of our data extraction and aggregation\cite{tgf2}. Fig.(\ref{video}) shows a snapshot of the video capturing part of a segment of the expressway of $\sim 50 m$, and we are able to accurately extract the position, velocity and acceleration of every vehicle appearing in the video. The resulting flow-density relation is also plotted, which shows very consistent behaviours from one day to another. The morning peak traffic has been captured, and since we are only interested in the congested traffic, we only analyse the vehicle dynamics during the peak hours, which is reflected by the red part of the plots in Fig.(\ref{video}).

\begin{figure}
\begin{center}
\includegraphics[height=12cm]{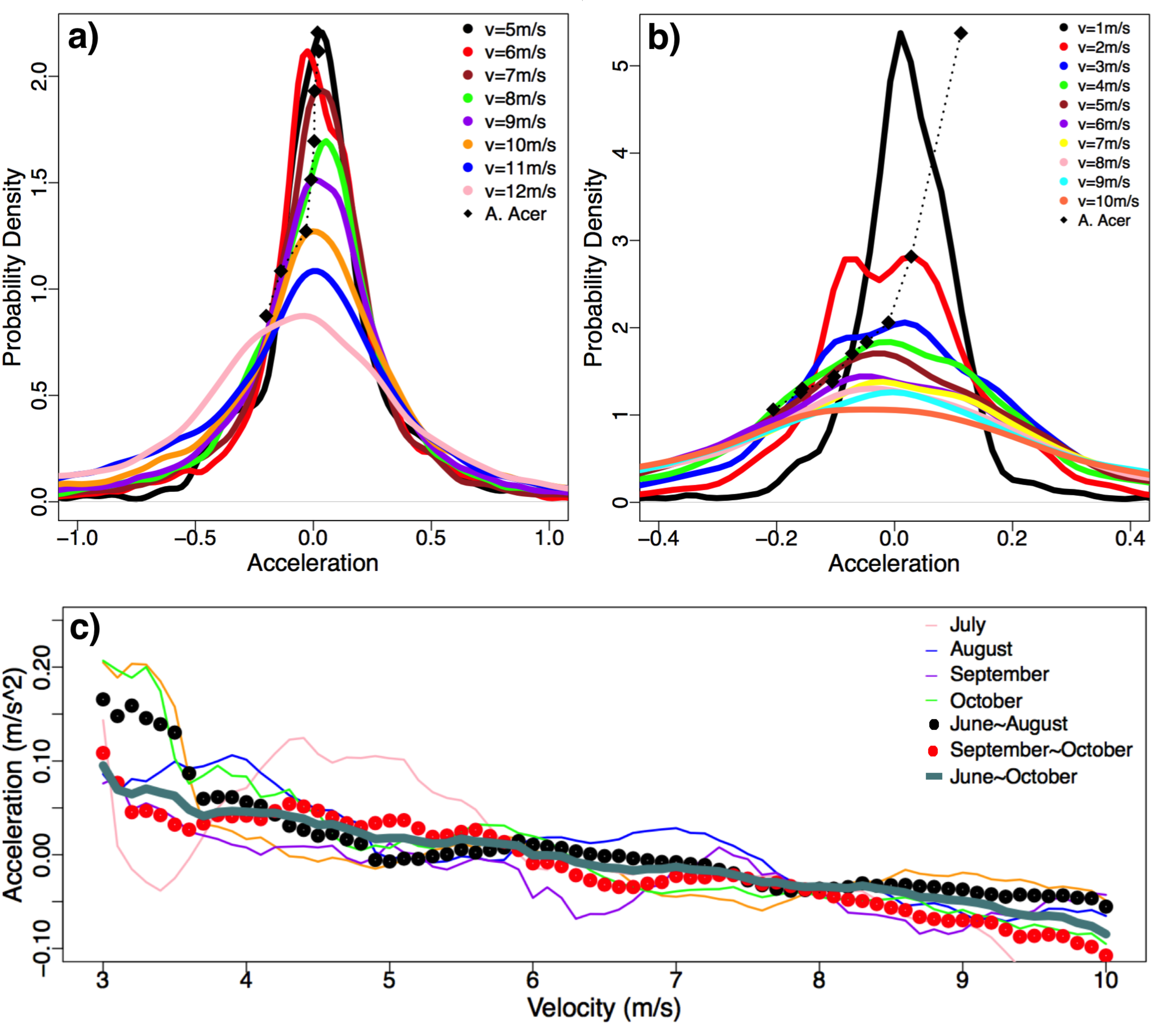}
\caption{The probability density distribution of the acceleration at $\Delta v=0$ and a). $h=11m$; b). $h=5m$. The black dots are the average acceleration at each velocity. The width of the density distribution generally increases as the velocity $v$ gets higher. c). The ``renormalised", or averaged acceleration at $\Delta v=0$ as a function of the velocity at $h=8m$. The average is done from data collected over one month, three months and six months. The fluctuation of the plot decreases with more data available. On average $3000$ vehicles are captured each day, and each vehicle yields around $200$ data points.}
\label{fluctuation}
\end{center}
\end{figure}

For each pair of vehicles passing through the video, we record the following vehicle's acceleration $a$, velocity $v$, bumper-to-bumper headway $h$ and the relative velocity $\Delta v$. The technical details of the data extraction can be found in\cite{tgf2}. In general, for the same $h,v$ and $\Delta v$ , different accelerations will be measured either from the same vehicle at different time/position, or from different vehicles. Such temporal and vehicle dependences are because of the many factors influencing the acceleration, and in 
Fig.(\ref{fluctuation}) we show examples of the distribution of the accelerations at $h=5m,11m$, with $\Delta v=0$ and different velocities.

The distribution profiles of the accelerations depend on the particular traffic system under study, and can be used to validate the stochastic terms in a stochastic microscopic or cellular automata model, though we will not study this issue here. We focus on obtaining the deterministic microscopic models with a representative model vehicle. Given the amount of data we have, we consider our traffic system to be fully sampled, and the ``renormalised" or effective model is obtained by averaging over all factors affecting the acceleration other than the dynamical parameters $h,v,\Delta v$. 

Formally, let the complete set of factors affecting the acceleration to be $\{h,v,\Delta v,\{f_i\}\}$, where $\{f_i\}$ contains factors such as vehicle index, following vehicles or vehicles further downstream, as well as other stochastic factors. Averaging over $\{f_i\}$ can be achieved as follows\cite{yangbo1}:
\begin{eqnarray}\label{renormalise}
\bar a\left(h,v,\Delta v\right)=\frac{1}{N_0}\sum_{\{f_i\}}a\left(h,v,\Delta v,\{f_i\}\right)
\end{eqnarray}
where $N_0$ is some proper normalisation factor. This can be precisely achieved by averaging over all the accelerations measured at the same $h,v,\Delta v$ from the video, with the proper sampling weight of $\{f_i\}$ intrinsically built in. Eq.(\ref{renormalise}) can be treated as the master microscopic traffic model that is deterministic with identical drivers, by simply letting $a_n=\bar a\left(h_n,v_n,\Delta v_n\right)$, where $n$ is the vehicle index increasing in the direction of the traffic flow and $\Delta v_n=v_{n+1}-v_n$.

If the traffic system is under-sampled, the resulting Eq.(\ref{renormalise}) will be fluctuating too much to be useful for model construction or the understanding of the traffic system. With more complete sampling, such fluctuations dampen to yield a smoother relationship between the renormalised acceleration and the dynamical parameters (see Fig.(\ref{fluctuation}) c). One should also be careful not to set the bin-size around $h,v,\Delta v$ to be too small. Not only does that reduce the sample size for a particular set of the dynamical parameters, there are also fundamental limitations on the resolution of headways and velocities from the driver perceptions. In this paper, the bin-size is $1m$ for headway and $1ms^{-1}$ for velocities, centered around the particular set of $h,v,\Delta v$. 

We will now present the results of the renormalised acceleration of the representative vehicle in the traffic system, as a function of its own velocity, the bumper-to-bumper headway to the preceding vehicle, and its relative velocity to the preceding vehicle. It is clear that the renormalised acceleration will only converge with enough data points within the bin of the same set of $h,v,\Delta v$. For the congested traffic, it is rare for two consecutive vehicles to be more than $20m$ apart. In general, it is also rare for two consecutive vehicles to have a velocity difference of more than $\pm 2ms^{-1}$. We are also careful to filter out manoeuvres involving lane changing. In general, we only plot the data at each $h,v,\Delta v$ when more than $5000$ data points are observed, so that the uncertainty in the averaged acceleration is less than $\pm 0.02 ms^{-2}$.

\begin{figure}
\begin{center}
\includegraphics[height=8cm]{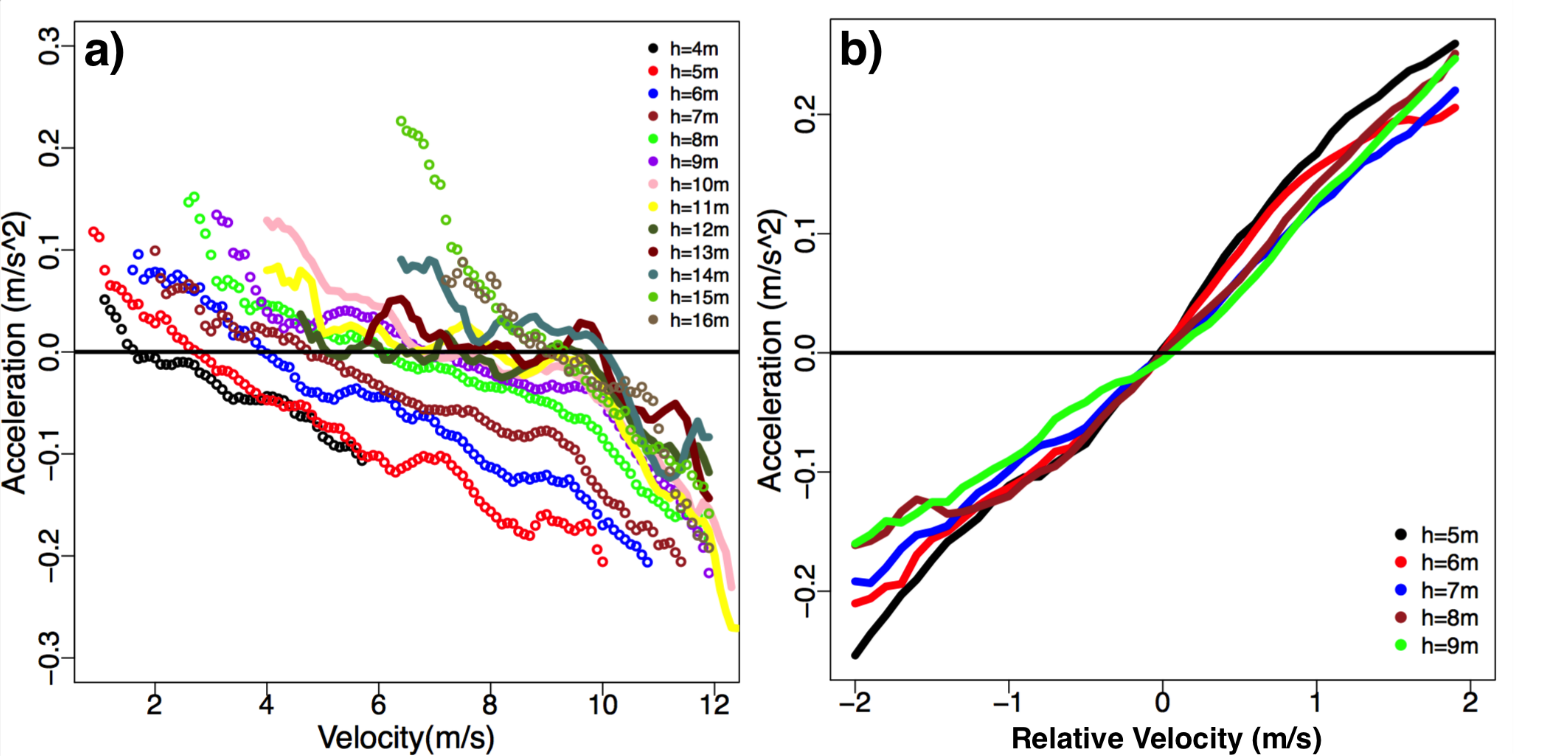}
\caption{a). Average acceleration as a function of velocity at different headway with zero relative velocity. The dotted plots intersect the x-axis only once, and the line plots have plateaus of near zero accelerations within a certain velocity range. b). Averaged acceleration as a function of $\Delta v$, when the velocity is equal to the optimal velocity.}
\label{profile}
\end{center}
\end{figure}

We first look at the cases where $\Delta v=0$. These cases are important in understanding the steady states of the traffic system, as well as in deciding if the system has a fundamental diagram (which is crucial for the three-phase traffic theory\cite{kernerbook}). The optimal velocity function (at least in the density range when the fundamental diagram does exist) can also be extracted by plotting the acceleration as a function of the velocity at different headways. In Fig.(\ref{profile}a) we can clearly see the general trend of the decrease in acceleration when the velocity of the vehicle increases. The interception with the x-axis gives the optimal velocity that generally increases with the headway as expected.

The most salient features of Fig.(\ref{profile}a) is the formation of plateaus close to zero acceleration, for the range of $10 m\le h\le14 m$, where the acceleration is very weakly dependent on the vehicle velocity. In that range, the drivers would rather just maintain the velocity as long as $\Delta v=0$, and their velocities are not too small or too large. The empirical data are reminiscent of the ``synchronisation gap" postulated by Kerner\cite{kernerbook}. It is now very clear that the multitudes of states with non-unique flow-density relationship does exist within a certain density and (density dependent) velocity range, independent of stochasticity or vehicle diversity. Due to the smallness of the accelerations, such states can last for a very long time and may even be treated as steady-states for all practical purposes. If such detailed driving behaviours are important for certain spatial-temporal patterns of the traffic flow, Fig.(\ref{profile}a) can lead to more accurate three-phase traffic models\cite{kernermodel1,kernermodel2}.

The dependence of the acceleration on the relative velocity is shown in Fig.(\ref{profile}b), when the vehicles are travelling at their optimal velocities. The gradients of the plots are symmetric about $\Delta v$ and only depends on the headways weakly, in contrast to the assumptions in some of the car-following models\cite{afvd}. When the vehicles are travelling at the optimal velocity (no acceleration if $\Delta v=0$), there is on average an equal tendency to accelerate or decelerate if the preceding vehicle starts pulling away or slowing down. 

We will now present a more quantitative analysis of the empirical data, by first noting that the microscopic master model can be expanded as follows\cite{yangbo2}:
\begin{eqnarray}
&&\bar a_n=\sum_{p,q}\lambda{p,q}\left( h_n\right)\left( \tilde v_n-V_{op}\left(h_n\right)\right)^p\Delta \tilde v_n^q\label{expansion1}\\
&&\lambda{p,q}\left(h_n\right)=\frac{1}{p!q!}\frac{\partial^{p+q}\bar a_n}{\partial^p \tilde v_n\partial^q\tilde\Delta v_n}\bigg|_{\begin{subarray}{l} \tilde v_n=V_{op}\left(h_n\right)\\\Delta \tilde v_n=0\end{subarray}}\label{expansion2}
\end{eqnarray}
Analyticity of $\bar a_n$ is assumed, as evidenced in Fig.(\ref{profile}) and argued in\cite{yangbo2}. The coefficients of expansion $\lambda_{p,q}$ have the dimension of the acceleration; $\tilde v_n=v_n/V_1\left(h_n\right), \Delta \tilde v_n=\Delta v_n/V_2\left(h_n\right)$, where $V_1, V_2$ are two headway dependent characteristic velocities obtained from the empirical data. We take $V_1$ to be on the order of the average velocity deviation from the optimal velocity, and $V_2$ that of the average $\Delta v$, respectively evaluated at each $h$. This makes sure all powers in Eq.(\ref{expansion1}) are smaller than $1$ in general, so that higher powers are less important if their coefficients are also small.

\begin{figure}
\begin{center}
\includegraphics[height=12cm]{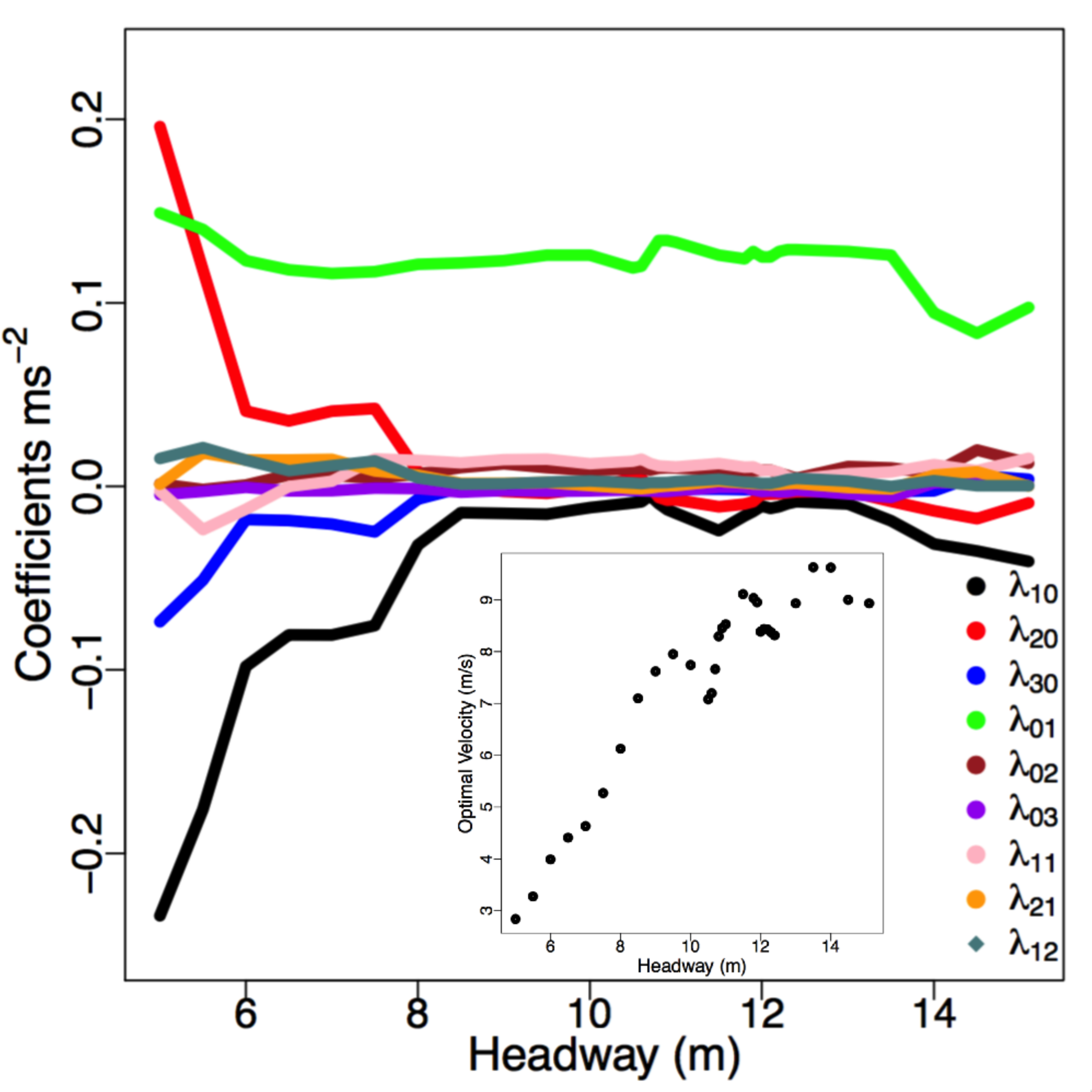}
\caption{Headway dependence of various coefficients of expansion. The inset is the dependence of the optimal velocity on the headway, where the fluctuation between $9m<h<14m$ is due to the plateau of almost zero accelerations. All plots are obtained from the fitting of the empirical data.}
\label{coefficients}
\end{center}
\end{figure}

The empirical data is fitted with a truncated master model by limiting $p,q\le 3$ in Eq.(\ref{expansion1}), using least-square non-linear regression. The results in Fig.(\ref{coefficients}) show the dominant terms are $\lambda_{01}$ and $\lambda_{10}$ of the lowest orders of expansion. There is a strong dependence of the coefficients on $h$, especially that of $\lambda_{10}$, explains some of the unrealistic features of the conventional optimal velocity model class\cite{afvd,bando,JiangR_PRE01}. Higher order coefficients are generally much smaller than the lower order coefficients, justifying our truncated model and also making it possible for simple effective models to be constructed from the master model.

For vehicles with small headways $h<8m$, the velocity dependence of the acceleration is strongly non-linear on the vehicle velocity. The prominence of $\lambda_{20}$ in this region shows drivers are more eager to accelerate than to decelerate, when their velocities deviate from the average optimal velocity. Such tendency seems to be reversed when the headways are large $h>14m$. In this region, drivers are much more cautious in accelerating even when their velocities are slower than what is perceived as safe. These can lead to interesting interpretation of the human driving psychology when the drivers are aware that the traffic is busy and congested during the peak hours.

The vanishing of the leading order of $\lambda_{10}$ as a function of $h$ has been predicted in\cite{yangbo1,yangbo2}, to explain the possible onset of the ``synchronised phase". In the region $9m<h<14m$, the only dominant coefficient is $\lambda_{01}$, indicating that the drivers are only sensitive to the relative velocity. However, if the velocities are too high or too low, acceleration will deviate from zero in a strongly non-linear way. 

From the modelling perspective, the range of $h>14m$ is relatively easier; only the linear terms with $\lambda_{10}$ and $\lambda_{01}$ are important, and the steady states have a unique flow-density relationship. The non-linear interactions between the vehicles are still strong, and depending on the detailed profile of the optimal velocity function, the traffic system can still be unstable against the formation of traffic jams, though headway fluctuations will generally develop when the coefficients are headway dependent. For densely populated roads where most of the headways are smaller than $8m$, the headway dependence of the coefficients will very likely lead to highly fluctuating traffic flows at high density (see Fig. 4e of \cite{yangbo2}), agreeing with the empirical observation\cite{kernerbook}. While this shows strong evidence of what can be captured by the proper deterministic microscopic models, further data collections for the full range optimal velocity function is needed for the complete determination of such models.

In conclusion, we show for the first time how the construction of the identical drivers for a traffic system can be validated with the properly averaged microscopic empirical data. We expect some of the important features from the expressway traffic system in Singapore to be universal. They particularly show strong empirical evidence that the non-linear interaction \emph{alone} is responsible for the multitude of steady or long lasting states covering a two-dimensional region of the flow-density plane. The collective sensitivity of the drivers on their own velocity and the relative velocity is also shown to be strongly headway dependent. Our observation setup is less suitable for characterising the free-flow traffic because of the relatively short segment of the road we are monitoring, leading to the scarcity of the relevant data. Free-flow traffic during off-peak hours are relatively easier to understand, though one should note that for the same $h,v,\Delta v$, the averaged acceleration in the free flow will be rather different, due to the different group of drivers and different driver expectations of the traffic conditions. While we did not focus on model calibration in this paper, our analysis can be useful for validating different assumptions employed in different traffic models, and for the construction of more realistic traffic models (deterministic or stochastic). It also serves as a good starting point to understand which detailed driving behaviours need to be modelled for the numerical simulations to be useful for practical applications.

\begin{acknowledgements}
This research was partially supported by Singapore A$^{\star}$STAR SERC ``Complex Systems" Research Programme grant 1224504056.  
\end{acknowledgements}

\end{document}